# Application of Machine Learning in Forecasting International Trade Trends

**Feras A. Batarseh**
Graduate School of Arts & Sciences - Data Analytics
Georgetown University
Washington, D.C. 20057
feras.batarseh@
georgetown.edu

**Munisamy Gopinath**
Department of Agricultural and Applied Economics
University of Georgia
Athens, Georgia 30602
m.gopinath@uga.edu

**Ganesh Nalluru**
Volgenau School of Engineering
George Mason University
Fairfax, Virginia 22030
gn@gmu.edu

**Jayson Beckman**
Economic Research Services
Department of Agriculture
Washington, D.C. 20024
jayson.beckman@usda.gov

**Abstract** – International trade policies have recently garnered attention for limiting cross-border exchange of essential goods (e.g. steel, aluminum, soybeans, and beef). Since trade critically affects employment and wages, predicting future patterns of trade is a high-priority for policy makers around the world. While traditional economic models aim to be reliable predictors, we consider the possibility that Machine Learning (ML) techniques allow for better predictions to inform policy decisions. Open-government data provide the fuel to power the algorithms that can explain and forecast trade flows to inform policies. Data collected in this article describe international trade transactions and commonly associated economic factors. Machine learning (ML) models deployed include: ARIMA, GBoosting, XGBoosting, and LightGBM for predicting future trade patterns, and K-Means clustering of countries according to economic factors. Unlike short-term and subjective (straight-line) projections and medium-term (aggregated) projections, ML methods provide a range of data-driven and interpretable projections for individual commodities. Models, their results, and policies are introduced and evaluated for prediction quality.

## Motivation and Background

In recent years, many countries are concerned about rising trade deficits (value of exports less imports) and their implications for employment and wages. For instance, the United States' goods and services trade deficit with China was $378.8 billion in 2018. Such numbers are forcing countries to either exit trade agreements or enforce tariffs, (e.g. Brexit, U.S. tariffs on Chinese goods). These shocks to global trade in commodities pose challenges to predict future trading patterns. In the United States, farm program costs, the President's Budget and recent compensation programs to address farmers' losses due to retaliatory tariffs depend on accurate trade predictions (from: USDA, 2018 – Trade damage estimation).

International economics has a long history of improving our understanding of factors causing trade and the consequences of free flow of goods and services across countries. Nonetheless, the recent shocks to the free-trade regime raise questions on the quality of earlier predictions and their applicability in the context of large trade disputes (Batarseh et al. 2018). To address these challenges, this article, identifies ML techniques appropriate for the international trade setting and tests their validity in making high quality projections. Recent technological advancements in ML as well as data democratization have also helped transparency, which is critical in the context of trade policy-making. Given the *Open Data* and *Big Data* initiatives presented in 2008 and 2012 (White House 2008), federal agencies are forced to share their data on public repositories such as www.data.gov. Econometric approaches identified have multiple ingredients that effect commodities' production and utilization, and hence, directly influence imports and exports of those commodities (Gevel et al. 2013). A field that can greatly aid with this analysis is Explainable AI (XAI) (Gunning 2019). XAI and ML can be instrumental in explaining previous and emerging patterns in data. This paper aims to address three main questions:

1- Do economic variables (such as GDP and population) associate with each other and a countries' exports?
2- Can boosting algorithms ensure learning and predictions from *country-commodity-year* cubical trade data?

3- Can ML techniques qualitatively improve the forecast from traditional econometrics?

We answer these questions in this article using experimental work. We also find that the ML data models developed here are scalable to all trade transactions, all over the world, and for all commodities.

## Related Work

Based on a recent study by the National Bureau of Economic Research (NBER), ML is **only recently** being applied to econometrics. ML has been applied across multiple domains; it has been employed in addressing challenges in healthcare (Reddy and Aggarwal 2015), education (Niemi et al.), and sports (Alamar 2013). To date, applications to studying trends in international trade are limited (except for a few narrow studies that are referenced in this paper). Storm et al. provided a comprehensive review of ML methods deployed to applied economics, especially their potential in informing policy decisions (Storm et al. 2019).

ML is commonly understood as a number of computational algorithms that extract hidden insights from large sets of data. In this study, multiple ML methods are applied to a big data set of international trade (imports/exports). A very distinct difference between ML and econometrics is that the latter aims to identify causalities, while the former is concerned with regressions, classifiers, clusters, associations and multiple other actionable outcomes.

Few studies have applied ML methods to economics. In 2013, Gevel et al. published a book called "the Nexus between Artificial Intelligence and Economics". It was one of the first few works that introduced agent-based computational economics (Gevel et al. 2013). One year later, Feng et al. studied economic growth in the Chinese province of Zhejiang using a *neural networks* model. Their method however, is very limited in scope, and proves difficult to deploy across other provinces in China or other geographical entities in other countries (Feng et al. 2014). Abadie et al. developed a similar model, but applied it to the rising tobacco economy in California (Abadie et al. 2010). In 2016, Milacic et al. expanded the scope, and developed a model for growth in GDP including its components: Agriculture, Manufacturing, Industry, and Services (Milacic et al. 2016). See also Kordanuli et al. for an application of neural networks for GDP predictions (Kordanuli et al. 2016). Falat et al. developed a set of ML models for describing economic patterns, but did not offer predictions (Falat et al. 2015). Experimental work presented in this paper utilizes ML methods in an optimized manner to provide predictions regarding trade of specific commodities and countries. Given the dimensions of the data, and the high number of variables involved, seven models (discussed in the *Methods* section) are developed and compared to explain international trade trends. The next section presents the datasets used, data collection, and data cleaning processes.

## Data Collection and Exploration

Data for this study has been collected from USDA's Foreign Agricultural Services' Global Agricultural Trade System (FAS - GATS) (USDA 2019). GATS is a system published by the United States Department of Agriculture. Additionally, economic data are collected from the World Bank's *World Integrated Trade Solution* (WITS 2019) and U.S. ITC's Gravity Portal (2019). The GATS system is used to assemble trade in seven major commodities: **Wheat**, **Milk**, **Rice**, **Corn**, **Beef**, **Soy**, and **Sugar**. Additionally, commodity data are merged with 30+ economic variables, such as: *Population*, *Currency*, *Island or Not*, *GDP of Origin (o)*, *GDP of Destination (d)*, *Distance*, *Landlocked or not*, *WTO Member*, *Hostility*, *EU Member,* and other ones (U.S. ITC Gravity Portal). Afterwards, the economic and commodity data are merged into a SQL database. An R code is used to merge on <u>country-to-country</u> trade transactions, as well as <u>year</u> of economic variables. The data are merged using an **Inner Join**. The 30+ economic variables' correlations are studied, results for the correlations are plotted in Figure 1.

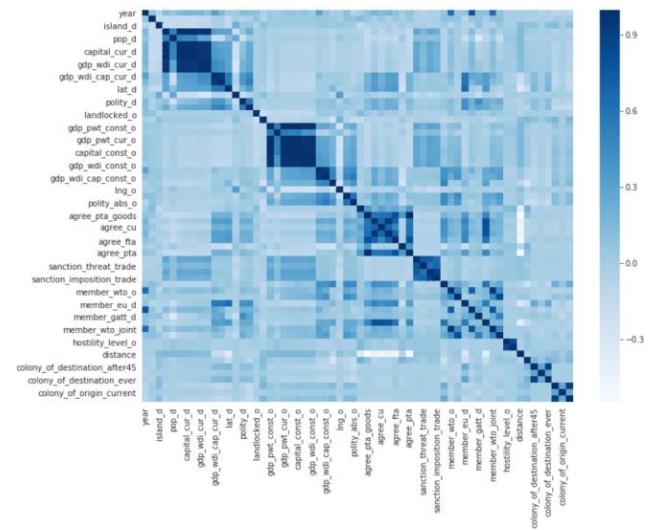

Figure 1: Correlations of 30+ Economic Variables

Highest economic correlations, for example, are found between: *population* and *whether the country is an island*, also, between *currency* and *GDP,* and between *WTO membership* and *Free Trade Agreements*; amongst other existing factors. Such insights support employing ML methods to assess if predictions better than traditional econometrics can be attained.

## ML Methods and their Results

In this study, supervised and unsupervised methods have been explored: Linear Regression, K-means clustering, Pearson correlations, Boosting, and Time Series such as Autoregressive Integrated Moving Average (ARIMA). Simple linear regression modeling is applied to the seven major commodities mentioned; our aim is to predict exports or imports of a specific commodity. For example, after importing required columns into a python environment, we deployed linear regression using the python libraries: *sklearn.linear_model* and *pandas* (Python 2019). An example is shown in Figure 2 of the top beef exporters of all time. Top countries exporting beef are: Australia, Germany, Netherlands, France, and United States.

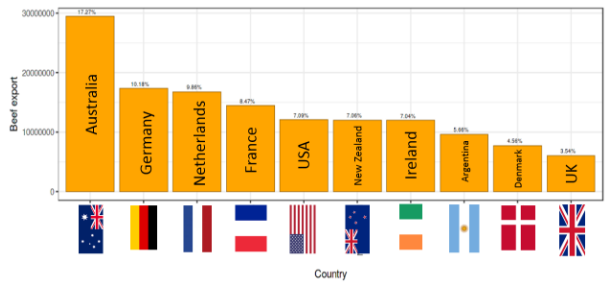

Figure 2: Top Beef Exporters

Data for beef trade are available from 1989 to 2018; years 2019-2021 are predicted (red line in Figure 3). As the figure illustrates, trade between nations is variant, and can change drastically over time; even for one commodity. Therefore, due to the high variance in the data, a simple regression model, although supervised, provides straight-line pointers to the future of beef trade (implying growth remains constant). Consequently, as an experimental model, we developed an unsupervised K-means clustering model to group countries into clusters (using *sklearn's cluster* and *K-Means* libraries in python). The clustering model yielded very expected results: China and the US ended up in Cluster 1, as the biggest exporters and importers. Cluster 2 has other major exporters and importers: Japan, Germany, Canada, UK, India, and France. Cluster 3 has less important importer countries, such as most third-world countries.

Subsequently, besides trade values, other economic variables are incrementally added to the modeling process. When all the economic variables are added, the aim is to identify which variables have the highest influence on trade predictions, and which ones could be controlled and tuned to change the forecasts. Different commodities had different rankings of economic variables, however, ***distance** (between the 2 countries undergoing trade)*, ***population of the exporter***, and ***GDP of both countries*** had the highest impact on whether two countries would trade one of the seven major commodities or not. Feature importance (Split of top economic variables) is illustrated in Figure 4. Consequently, ARIMA is applied to beef trade, the advantage of ARIMA is that it provides univariate predictions that improve the output. ARIMA results are presented in Table 1; it illustrates the high and low confidence intervals of the model.

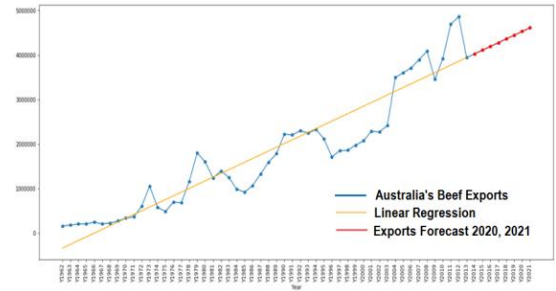

Figure 3: Australia's Beef Exports 1988-2021

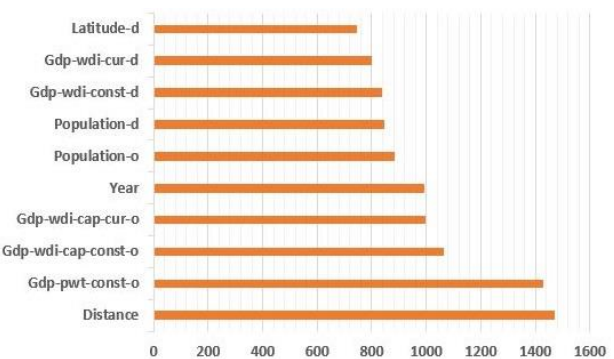

Figure 4: Variable Importance in Predicting Trade

Table 1: ARIMA Forecasting of Beef Trade Trends

| Year | Actual | Forecast | Low 80 | High 80 | Low 95 | High 95 |
|---|---|---|---|---|---|---|
| 2014 | 6939233 | 5594243 | 5208462 | 5980024 | 5004242 | 6184244 |
| 2015 | 7357932 | 5698666 | 5153090 | 6244242 | 4864279 | 6533053 |
| 2016 | 5921218 | 5803089 | 5134898 | 6471281 | 4781178 | 6825000 |
| 2017 | 5843209 | 5907513 | 5135951 | 6679074 | 4727511 | 7087514 |
| 2018 | X | 6011936 | 5149304 | 6874567 | 4692654 | 7331217 |
| 2019 | X | 6116359 | 5171393 | 7061325 | 4671159 | 7561559 |
| 2020 | X | 6220782 | 5200102 | 7241462 | 4659787 | 7781777 |
| 2021 | X | 6325205 | 5234053 | 7416358 | 4656432 | 7993979 |

Afterwards, boosting has been applied to elevate the quality of the models. Three different boosting models are deployed: Gradient Boost (GBoost), Extreme Gradient Boosting (XGBoost), and Light Gradient Boosting Decision Tree (LightGBM); their results are presented next.

## Boosting Hyper-Parameter Settings

Python libraries are used to deploy the boosting trees (GBoost, XGBoost, and LightGBM) (Ke et al. 2017). After multiple iterations and hyper-parameters' tuning, LightGBM performed best for most commodities. A boost-ing algorithm is an algorithm that converts weak learners to strong learners. It is a method that improves predictions' quality of a model (through $R^2$ measures). Boosting trains weak learners sequentially, and in every cycle, each trying to correct its predecessor. Results for trade predictions through the XGBoost Model scored predictions' quality = **69%**, and through LightGBM scored a quality of **88%** (in contrast, GBoost scored the lowest of the three approaches). Parameter tuning for boosting models include: Number of leaves, Maximum Depth of the tree, Learning Rate, and Feature fraction. Small learning rates are optimal (0.01), with large tree depths. Additionally, to speed up training and avoid over-fitting, feature fraction is set to 0.6; that is, selecting 60% of the features before training each tree. Early stopping round is set to 500; that allowed the model to train until the _validation score_ stops improving. Maximum tree depth is set to 8. Those settings led to the best output through LightGBM. *Sugar* for instance had an $R^2$ score of **0.73**, **0.88** for *Beef*, and **0.66** for *Corn*. These initial results confirm the applicability of ML methods to projecting trade patterns and also point to accuracy gains over traditional approaches.

## Conclusions

This article introduced ML models for international trade settings and posed questions on their applicability and prediction quality. Methods presented in this paper allowed for the extractions of the best economic variables that would affect trade of specific commodities. As mentioned for beef for example, *distance* had the highest effect (i.e. the US is better off trading beef to Canada and Mexico, its two closest neighbors). While Australia, being an island, has to focus its policies for beef exports on GDP measures, and the population of the importer. Feature Importance for all economic variables are *(name: split, gain.)*: *Distance: 1469, 6.38. GDP of Exporter: 1431, 6.22. Year: 993, 4.318. Population of Exporter: 882, 3.83. Population of Importer: 847, 3.68. Currency of Importer: 801, 3.48.*

The experimental work in this article indicates the high relevance of ML for predicting a range of trade patterns with a greater accuracy than traditional approaches. For example: Over 2006-17, the USDA forecast accuracy (*Ag outlook reports*) was less than 35% (USDA, 2018). Their accuracy improves to 92% *only* after having actual data for *three-quarters of the forecasted year*. Models presented in this paper offer forecast accuracy in the 69% – 88% range. Our models also offer an alternative to econometric approaches, which are seldom cross-validated. We offer a superior alternative to current approaches in public sector forecasting of agricultural trade flows. We rely on data – instead of complex behavioral models with assumptions solved by accessing information from a myriad of studies – and deep learning from data to allow for alternative and robust specifications of complex economic relationships. ML models allow for simulation of trade outcomes in alternative policy scenarios. This work is to be expanded to more commodities and models, specifically ones that produce interpretable results for policy makers.